\documentclass{iau} 
\usepackage{graphicx}

\title[The \helmod{} Model for Heliopheric propagation] 
{The \helmod{} Monte Carlo Model for the Propagation of Cosmic Rays in Heliosphere }

 \author[Boschini et al.]   
{{Matteo~J.} Boschini$^{1,3}$, 
 Stefano {Della~Torre}$^1$,
 Massimo Gervasi$^{1,2}$, 
 Davide Grandi$^{1,2}$,  
 Giuseppe {La~Vacca}$^{1,2}$, 
 Simonetta Pensotti$^{1,2}$, 
 {Pier~Giorgio} Rancoita$^1$, 
 Davide Rozza$^{1,2}$ 
 \and Mauro Tacconi$^{1,2}$}
 
 \affiliation{$^1$INFN Sezione di Milano Bicocca, I-20126 Milano, Italy\\ email: {\tt stefano.dellatorre@mib.infn.it} \\[\affilskip]
$^2$Universit\`a degli Studi di Milano Bicocca, I-20126 Milano, Italy\\
$^3$Cineca, Segrate, Italy}
 
\pubyear{2017}
\volume{335}  
\jname{Space Weather of the Heliosphere: Processes and Forecasts}
\editors{C. Foullon \&	O. Malandraki, eds.}

\usepackage{amsmath}
\newcommand{\galprop}{GALPROP}
\newcommand{\helmod}{\textsc{HelMod}}

\begin{document}

\maketitle

\begin{abstract}
The heliospheric modulation model HelMod solve the transport-equation for the Galactic Cosmic Rays propagation through the heliosphere down to Earth. 
It is based on a 2-D Monte Carlo approach, that includes a general description of the symmetric and antisymmetric parts of the diffusion tensor, thus, properly treating the particle drift effects as well as convection within the solar wind and adiabatic energy loss. 
The model was tuned in order to fit 1) the data observed outside the ecliptic plane at several distances from the Earth and 2) the spectra observed near the Earth for both, high and low solar activity periods. Great importance was given to description of polar regions of the heliosphere. We present  the flux for protons, antiprotons and helium nuclei computed along the solar cycle 23-24 in comparison with experimental observations and prediction for the full solar cycle 24. 
\end{abstract}

\firstsection 
\section{Introduction}
The intensity of Galactic Cosmic Rays (GCRs) observed at Earth changes with time according to solar activity.
The overall effect of heliospheric propagation on the spectral flux of GCRs is called \textit{solar modulation}.
In this paper we present the \helmod{} Model for the description of the solar modulation, as reported by~\cite{Boschini2017AdvSR}, and its web interface. A stand-alone python module, fully compatible with GalProp, was developed for the calculation of solar modulation effects, resulting in a newly suggested set of local interstellar spectra as reported by~\cite{2017ApJ...840..115B}. 
Finally we present forecasts for GCRs intensities of Protons and Helium nuclei up to the next solar minimum that we estimate to occour in 2020.

\section{\helmod{} Model for GCR Propagation in the Heliosphere\label{Sect::Helmod}}
GCRs propagation in the heliosphere was first studied by \cite{1965P&SS...13....9P}, who formalized the transport equation, also referred to as Parker equation (see, e.g., discussion in~\cite[Bobik et al. 2012]{Bobik2011ApJ}).
%
The Parker equation can be solved by means of numerical techniques. The \helmod{} code computes the CR propagation from the Termination Shock (TS) down to the Earth orbit using a Monte Carlo approach
(for details, see description in~\cite[Bobik et al. 2016]{BobikEtAl2016}).
The present form of diffusion parameter (as  defined in \cite[Boschini at al. 2017a]{Boschini2017AdvSR}) include a scale correction factor related to drift contribution. As discussed by~\cite{2017ApJ...840..115B} this correction is evaluated for protons during positive HMF polarity period to account the presence of latitudinal structure in the Galactic CRs spatial distribution. More details on the \helmod{} model and its implementation can be found in~\cite{Bobik2011ApJ},~\cite{DellaTorre2013AdvAstro} and~\cite{Boschini2017AdvSR}.
\par
As discussed by~\cite{Boschini2017AdvSR}, the validity of the \helmod{} Model is verified down to 1 GV in Rigidity (1 GeV in equivalent Kinetic energy for electrons). Lower energies are not considered in the present work since they need further improvement in the description of solar modulation in the outer heliosphere -- from TS up to interstellar space -- as well as the inclusion of turbulence in drift coefficient evaluation.

\subsection{Local Interstellar Spectrum}\label{Sect::LIS}
\begin{figure}[tbh]
\centerline{
\includegraphics[width=0.55\textwidth]{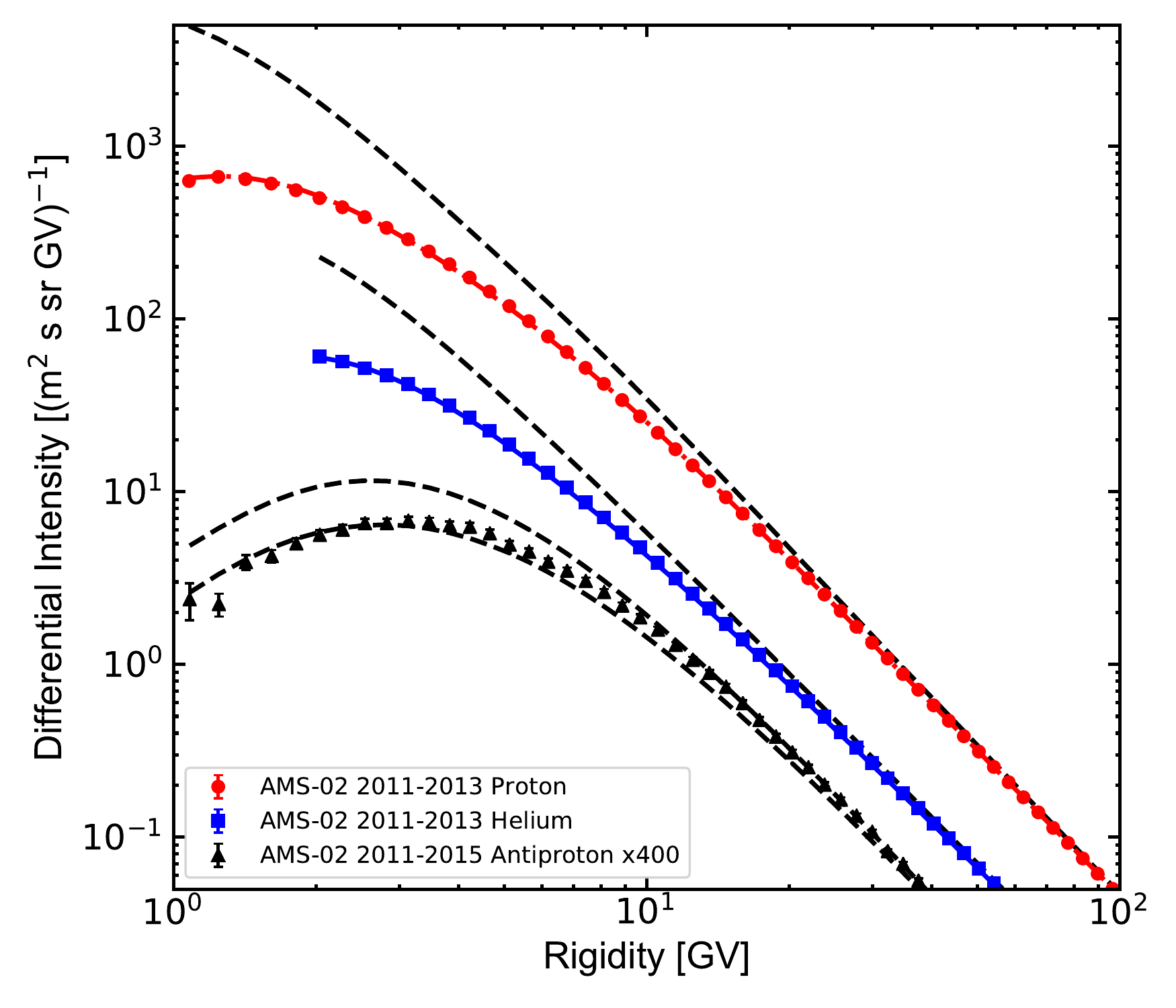}
}
\caption{Differential intensity of galactic protons (red points), helium nuclei (blue squares) and Antiprotons (black triangle)  measured by AMS-02 compared with modulated spectra from \helmod{}. Antiproton spectrum is scaled multiplying by a factor 100. Black dashed lines are the \galprop{} LIS’s (see text).\label{fig:HelModMod}
}
\end{figure}

The local interstellar spectrum (LIS) is assumed to be isotropic along the heliosphere boundary. 
Nowadays LIS parametrization is constrained  by measurements mainly from Voyager probes, at low energy, and AMS-02, at high energy. As described by~\cite{2017ApJ...840..115B}, in order to derive the physically motivated LIS of CR species, an iterative procedure was
developed to feed the \galprop{} output into \helmod{} to compare with AMS-02 data as observational constraints.
A Markov Chain Monte Carlo (MCMC) interface to  GALPROP was developed and described in~\cite{2017ApJ...840..115B}.
Main propagation parameters were left free in the scan using the 2D GALPROP model (\cite[Masi 2016]{Masi2016}). 
Parameters of the injection spectra, such as spectral indices and the break rigidities, were also left free, but their exact values depend on the solar modulation. As a matter of fact, the low energy part of the spectra are tuned together with the solar modulation parameters.
In the scan we used all published AMS-02 data on protons~(\cite[Aguilar et al. 2015a]{2015a_AMS}), helium~(\cite[Aguilar et al. 2015b]{2015b_AMS}), B/C ratio~(\cite[Aguilar et al. 2016a]{2016a_AMS}) and electrons~(\cite[Aguilar et al. 2014]{2014_AMS}), while antiproton data~(\cite[Aguilar et al. 2016b]{2016b_AMS}) were explicitly excluded in order to make a prediction of the antiproton spectrum based on GCR primaries. 
The so obtained set of LISs was presented in~\cite{2017ApJ...840..115B} and reported in Fig.~\ref{fig:HelModMod}.  In Fig.~\ref{fig:HelModMod} the differential intensity of galactic protons, helium nuclei and antiprotons, observed by AMS-02, are compared with spectra modulated by \helmod{}.


\begin{figure}[bth]
\centerline{
\includegraphics[width=0.98\textwidth]{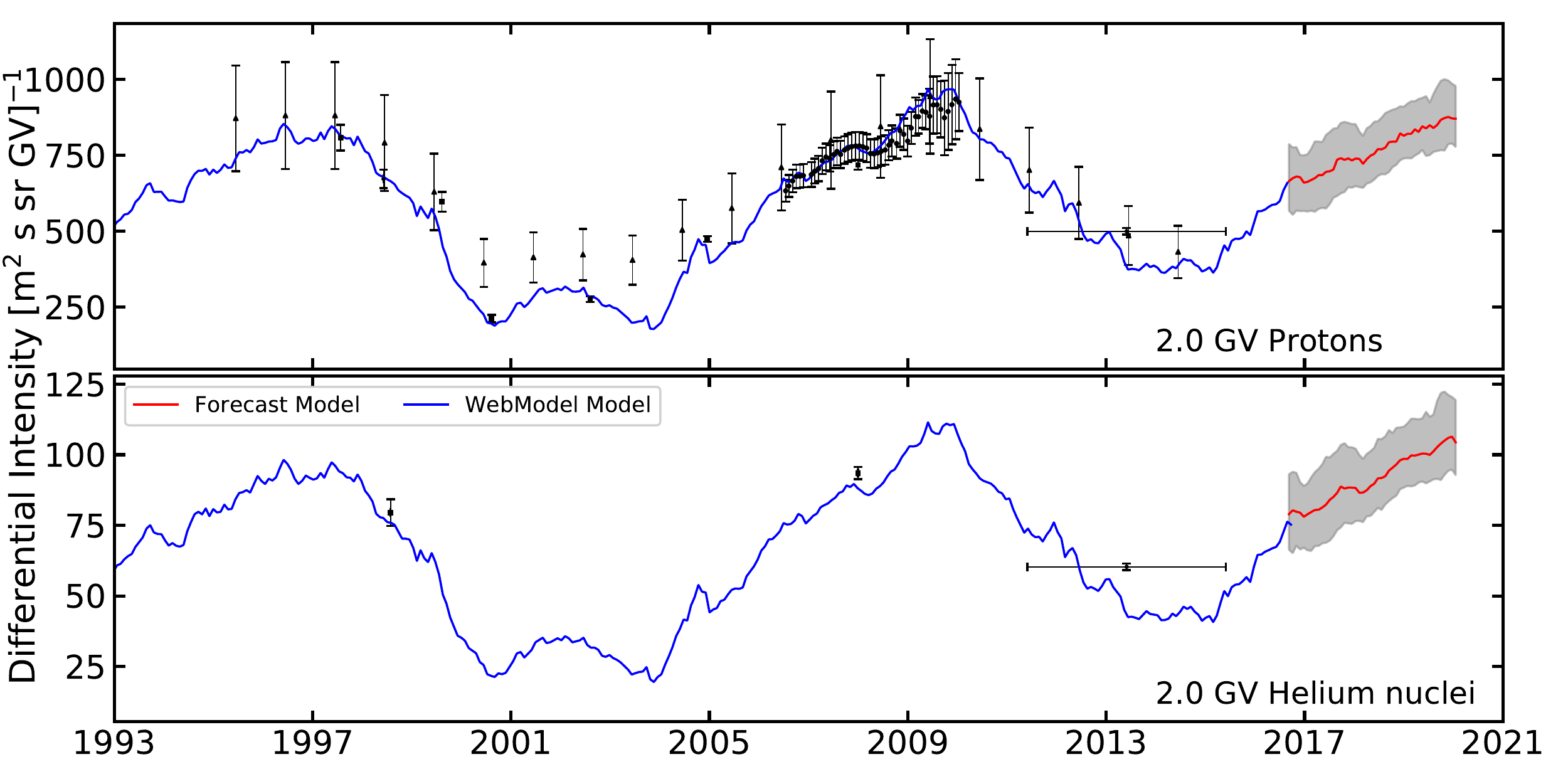}
}
\caption{Differential intensity of galactic proton  and Helium nuclei as a function of time. Experimental points are described in the text. Blue solid lines are obtained by \helmod{} on-line solar modulator. Red solid line are the forecast of protons and helium nuclei up to the end of 2020; shaded bands represents 20\% uncertainty.\label{fig:TimeVariation}
}
\end{figure}

\section{Long term solar modulation and forecasts}\label{Sect::TimeEvolution}

To simplify the calculations, we developed a python script reading the \galprop{} output and providing the modulated spectrum for periods of selected experiments. The module, as described in Section 3.1 of~\cite{2017ApJ...840..115B}, uses pre-evaluated normalized probability functions from \helmod{} thus this method dramatically accelerates the modulation calculations while maintaining the accuracy of the full-scale simulation.

The \helmod{} python module can be downloaded from a dedicated website or used on-line\footnote{\helmod{} (version 3.4), 2017, http://www.helmod.org.}. The LISs showed in Fig.~\ref{fig:HelModMod} are provided as default choice, however the user can also specify his own LIS in \galprop{} output or plain-text file formats. The on-line interface allows a direct  comparison with selected experimental data sets or provide modulated spectra for each Carrington Rotations from 1996 up to 2016. In Fig.~\ref{fig:TimeVariation} we show the differential intensity of protons and helium nuclei GCRs at 2 GV as a function of time (blue solid line) computed by means of on-line solar modulator. Simulations are compared with experimental observations at same rigidity provided by SOHO/EPHIN (black triangle,~\cite[K{\"u}hl et al., 2016]{2016SoPh..291..965K}), BESS (black squares,~\cite[Shikaze et~al. 2007]{bess_prot},~\cite[Abe et al., 2016]{BESS2007_Abe_2016}), PAMELA (black points,~\cite[Adriani et~al. 2013]{PamelaProt2013}) 
and 
AMS (black stars,~\cite[Aguilar et~al. 2002]{AMS01_prot}~\cite[Aguilar et al. 2015a]{2015a_AMS},~\cite[Aguilar et al. 2015b]{2015b_AMS}). Simulations well agree with experimental data within one standard deviation (two standard deviations at solar maximum for SOHO/EPHIN observations only).
\subsection{Solar Modulation forecast}
\helmod{} is also able to provide forecast of GCRs intensity. Using a procedure derived from~\cite{2011GeoRL..3819106O} and \cite{2012GeoRL..3919102O}, we are able to forecast the number of smoothed sunspot number (SSN), tilt angle of neutral sheet, magnetic field at Earth and Solar Wind velocity in forthcoming years. 
From the analysis of previous 10 solar cycles we derived the average variation of parameters along an ideal solar cycle of fixed average duration. 
The length and amplitude of the variation are then rescaled in order to fit measured values of SSN during Solar Cycle 24. The analysis found a cycle duration that is consistent with 11 years, thus, since the beginning of Solar Cycle 24 is set to be in 2009, the next solar minimum is expected to occur at the end of 2020. 
In Fig.~\ref{fig:TimeVariation} we report in red the forecast of protons and helium nuclei intensity from end 2016 to 2020. 
The shaded bands represent the uncertainties, i.e. $\sim20\%$, that is dominated by uncertainties on sunspot number forecast.

\section{Conclusion}
We presented the 2-D Monte Carlo Heliospheric Modulation Model, i.e. \helmod{}, to evaluate GCR modulated spectra at Earth.
The model includes details on the individual processes occurring in the heliosphere particle propagation, thus providing a comprehensive description of the solar modulation. The model is tuned by using several proton observations at Earth in different solar activity periods, treating in a unified description the high and low solar activity periods. Moreover \helmod{}, jointly with \galprop{} and thanks to the precise observations of AMS-02, provided unprecedentedly tight constraints to proton and Helium LISs.
To make the calculation available to the community, a python module and a dedicated website are developed to provide modulated spectra for selected GCR experiments in different solar activity periods. The model now includes also the forecast of future GCRs intensity until the end of cycle 24. The model estimate the next solar minimum will occur in 2020.

\bigskip 
\begin{acknowledgments}
We wish to specially thank Pavol Bobik, Giuliano Boella, Karel Kudela, Marian Putis and Mario Zannoni for their support to the \helmod{} code and many useful suggestions.
This work is supported by ASI (Agenzia Spaziale Italiana) under contract ASI-INFN I/002/13/0 and ESA (European Space Agency) contract 4000116146/16/NL/HK. 
\end{acknowledgments}

\end{document}